\documentclass[twocolumn,amsmath,showkeys,showpacs,prb]{revtex4-1}
\usepackage{dcolumn}
\usepackage{bm}
\usepackage[T1]{fontenc}
\usepackage{amsmath}
\usepackage{graphicx}
\usepackage{tabularx}
\usepackage{subfigure}

\usepackage{hyperref}
\hypersetup{colorlinks=true, linkcolor=blue, citecolor=blue, urlcolor=blue, pdftitle={Structural and Non-collinear Magnetism in Predicted Post-perovskite NaMnF$_3$ Fluoride  From Density Functional Theory}, pdfauthor={A.C. Garcia-Castro}}

\begin{document}
\title{First-Principles study of vibrational and non-collinear magnetic properties of the perovskite to postperovskite pressure transition of NaMnF$_3$}
\author{A. C. Garcia-Castro$^{1,2}$,  A. H. Romero$^{3,2}$ and E. Bousquet$^{1}$}
\affiliation{$^1$Physique Th\'eorique des Mat\'eriaux, Universit\'e de Li\`ege, B-4000 Sart-Tilman, Belgium}
\affiliation{$^2$Centro de Investigaci\'on y Estudios Avanzados del IPN, MX-76230, Quer\'etaro, M\'exico.}
\affiliation{$^3$Physics Department, West Virginia University, WV-26506-6315, Morgantown, USA.}

\begin{abstract}
We performed a first-prinicples study of the structural, vibrational, electronic and magnetic properties of NaMnF$_3$ under applied isotropic pressure. 
We found that NaMnF$_3$ undergoes a reconstructive phase transition at 8 GPa from the $Pnma$ distorted perovskite structure toward the $Cmcm$ post-perovskite structure.
This is confirmed by a sudden change of the Mn--F--Mn bondings where the crystal goes from corner shared octahedra in the $Pnma$ phase to edge shared octahedra in the $Cmcm$ phase.
The magnetic ordering also changes from a $G$-type antiferromagnetic ordering in the $Pnma$ phase to a $C$-type antiferromagnetic ordering in the $Cmcm$ phase.
Interestingly, we found that the high-spin $d$-orbital filling is kept at the phase transition which has never been observed in the known magnetic post-perovskite structures.
We also found a highly non-collinear magnetic ordering in the $Cmcm$ post-perovskite phase that drives a large ferromagnetic canting of the spins. 
We discuss the validity of these results with respect to the $U$ and $J$ parameter of the GGA$+U$ exchange correlation functional used in our study and conclude that large spin canting is a promising property of the post-perovskite fluoride compounds.
\end{abstract}


\maketitle

\subsection{Introduction}
Perovskite materials with stoichiometry $A$BX$_3$ are well known in material sciences due to their numerous properties of high interest for technological applications.  
These properties are very diverse such as ferroelectricity,\cite{Cohen1992} ferromagnetism\cite{Khomskii-2001}, piezoelectricity\cite{Cohen1998PRL}, pyroelectricity\cite{morozovska2012} or superconductivity.\cite{Rao1990} 
Going further, recent researches focused on the achievement of multifunctionalities in a single material that would open the door to new types of applications such as computer memories or spintronic.\cite{Fabian2004, spintronic-bibes2008}  
One of the main features of these compounds is that their properties can be conveniently tuned by external control parameters such as electric and magnetic fields, temperature, strain or pressure.\cite{Zubko2011,Chu2007} 
Hydrostatic pressure appeared to be an interesting external parameter for the understanding of the perovskite physics since they exhibit profuse structural, electronic and magnetic phase transitions.\cite{RAngel2005}  
A particularly appealing pressure phase transition is the perovskite to post-perovskite ($pPv$) phase transition that has been observed in MgSiO$_3$ crystal in 2004.\cite{Murakami2004,Tsuchiya2004}
The discovery of this $pPv$ phase transition had important consequences in the physics of the earth mantle where MgSiO$_3$ is abundant and is under very high pressure conditions.\cite{Murakami2004,Tsuchiya2004}
The $pPv$ transition has then attracted great interests and has been observed in other oxide and fluoride perovskites such as MgGeO$_3$\cite{Hirose2005}, CaSnO$_3$\cite{Tateno2005}, NaMgF$_3$\cite{Martin2006a} or NaCoF$_3$\cite{Yusa2012,Dobson2011} (see Table \ref{tab:Cmcm-all}).
Interestingly, the fluoroperovskite crystals have a lower pressure and temperature $pPv$ phase transition than the oxides and thus making them attractive for the study of the $pPv$ phase.
Furthermore, it has been reported the possibility to keep the $pPv$ phase at atmospheric pressure when releasing the pressure in fluroroperovskites \cite{Shirako2012} while all the oxides return to their $Pnma$ distorted perovskite ground state.\cite{Yusa2012}
Besides the interest for the $pPv$ transition in earth mantle studies, it has been recently highlighted the existence of a ferroelectric instability in the cubic phase of $A$BF$_3$ perovskites that originates from a geometric effect rather than a dynamical charge transfer as it the case in oxides.\cite{acgarciacastro2014}
It has also been shown the possibility to induce a ferroelectric polarization in NaMnF$_3$ when applying an epitaxial constraint which leads to new possibilities to make multiferroic materials within the family of fluoroperovskites.
The study of the fluoroperovskites under pressure is then of high interest to scrutinize the reasons and origins of their unique properties and their potential multiferroic character.

\begin{table}[htbp!]
\caption{Lattice parameters (\AA) of the $pPv$ phase and related temperature (T, in Kelvin) and pressure (P, in GPa) transitions from the $Pnma$ phase to the $pPv$ phase of fluroride and oxide perovskites. 
In the $pPv$ phase, the $A$ and $B$ cations occupy the $4a$ and $4c$ Wyckoff positions and the O/F anions occupy the $4c$ and $8d$ positions. 
Most of the oxides have higher pressure and temperature transitions than the fluorides.
**This work.}
\centering
\begin{tabularx}{\columnwidth}{X c c c c}
\hline
\hline 
Crystal    & $a/b/c$   & P &  T & Ref  \rule[-1ex]{0pt}{3.5ex} \\
\hline
                  \multicolumn{5}{c}{Oxides $AB$O$_3$}          \rule[-1ex]{0pt}{3.5ex} \\
\hline
MgSiO$_3$     &      2.456/8.042/6.093      &   121    &      300    & \cite{Murakami2004}  \rule[-1ex]{0pt}{3.5ex}\\
MgGeO$_3$    &      2.613/8.473/6.443      &   78    &      300    & \cite{Hirose2005}  \rule[-1ex]{0pt}{3.5ex}\\
MnGeO$_3$    &      2.703/8.921/6.668      &   57 - 65    &      1200 - 2400    & \cite{Tateno2005}  \rule[-1ex]{0pt}{3.5ex}\\
CaSnO$_3$    &      2.854/9.343/7.090      &   56    &      300    & \cite{Tateno2005}  \rule[-1ex]{0pt}{3.5ex}\\
CaIrO$_3$       &      3.145/9.862/7.298      &   1 - 3    &      1648 - 1798    & \cite{Bogdanov2012, Ballaran2007}  \rule[-1ex]{0pt}{3.5ex}\\
CaRuO$_3$    &      3.115/9.827/7.296      &   21 - 25    &      1173 - 1373    & \cite{Shirako2010, Kojitani2007, Shirako2011}  \rule[-1ex]{0pt}{3.5ex}\\
CaPtO$_3$     &      3.126/9.920/7.351      &   4    &      1073    & \cite{Ohgushi2008, Inaguma2008}  \rule[-1ex]{0pt}{3.5ex}\\
CaRhO$_3$    &      3.101/9.856/7.264      &   12 - 27    &      1273 - 1673    & \cite{Shirako2009, Yamaura2009}  \rule[-1ex]{0pt}{3.5ex}\\
NaIrO$_3$      &      3.040/10.358/7.177    &   4 - 5    &      1073    & \cite{Nairo2011}  \rule[-1ex]{0pt}{3.5ex}\\
\hline
                            \multicolumn{5}{c}{Fluorides $AB$F$_3$}           \rule[-1ex]{0pt}{3.5ex} \\
\hline
NaMgF$_3$    &      2.716/8.381/6.849      &   18 - 30    &      300 -   & \cite{Martin2006a}  \rule[-1ex]{0pt}{3.5ex}\\
NaZnF$_3$    &      3.034/10.032/7.450      &   13 - 18    &      300 -   & \cite{Yakovlev2009}  \rule[-1ex]{0pt}{3.5ex}\\
NaCoF$_3$    &      3.064/10.123/7.468      &   11 - 14    &      300 -   & \cite{Yusa2012, Dobson2011}  \rule[-1ex]{0pt}{3.5ex}\\
NaNiF$_3$    &      3.026/10.058/7.401      &   16 - 18    &      1273 - 1473    & \cite{Shirako2012, Dobson2011}  \rule[-1ex]{0pt}{3.5ex}\\
NaMnF$_3$    &     3.042/9.839/7.416      &   8    &      0    & **  \rule[-1ex]{0pt}{3.5ex}\\
\hline
\hline
\end{tabularx}
\label{tab:Cmcm-all}
\end{table}

In this paper, we propose to study from first-principles calculations the structural, electronic and magnetic properties of NaMnF$_3$ as function of applied hydrostatic pressure. 
The results are presented as follows.
We first determine the structural ground state structures over a wide range of pressure conditions and we predict a transition from the $Pnma$ ground state to a   $pPv$ phase that has not been explored experimentally.
We analyse the $pPv$ phase transition through a vibrational characterization (phonons) and a structural study.
We then focus on the analysis of the magnetic behaviour with a special interest in the non-collinear magnetic ground states allowed in the $pPv$. 
We found a high-spin ground state associated with a large ferromagnetic canting that has not been reported in any of the previous $pPv$ transition observed in other compounds.
The validity of the results are discussed with respect to the values of the DFT$+U$ $U$ and $J$ parameters.

\subsection{Computational details}
Total energy calculations were performed with the Vienna Ab-initio Simulation Package (VASP) \cite{Kresse1996,Kresse1999} and within the projected augmented wave (PAW) method to describe the valence and core electrons.\cite{Blochl1994} 
The electronic configurations taken into account in the PAW pseudopotentials are as follows: seven valence electrons for Na (2p$^6$3s$^1$), thirteen for Mn (3p$^6$4s$^2$3d$^5$) and seven for F (2s$^2$2p$^5$). 
We used the Generalized Gradient Approximation (GGA) exchange correlation functional within its PBEsol variant~\cite{Perdew2008} and we corrected it to increase the \emph{d}-electron localization by means of the DFT+$U$ approximation, with $U$ = 4.0 eV and $J$ = 0 of the Liechtenstein formalism \cite{Liechtenstein1995}.
Due to the magnetic character of NaMnF$_3$, the spin degree of freedom have been included in the calculations. 
Non-collinear magnetism calculations have been carried out by including the spin-orbit coupling as implemented in VASP by Hobbs \emph{et. al.}\cite{Hobbs2000}.
Different values of the $U$ and $J$ parameters were systematically varied to study the influence of these parameters on the results.
The reciprocal space has been discretized by a Monkhorst-Pack \emph{k}-point mesh of (6$\times$4$\times$6) and the plane wave expansion has been limited by an energy cut-off of 800 eV.
These convergence parameters were necessary in order to have a resolution on the force of less than 1$\times$10$^{-4}$ eV/\r{A} and a resolution on the energy of 0.01 meV. 
Analysis of the optimal structure as a function of pressure was carried out according to the third-order Birch-Murnaghan equation of state (BM-EOS) \cite{Birch1947} by fitting the total energy dependence of the primitive cell volume per molecule. 
Vibrational properties were computed through the formalism of density functional perturbation theory (DFPT)\cite{Gonze1995-1, Gonze1995-2}  as implemented in  VASP and post-processed with the Phonopy code \cite{Togo2008}.

\subsection{Structural and vibrational properties of the $Pnma$ phase of NaMnF$_3$}
At room temperature and atmospheric pressure, the sodium manganese fluoride compound NaMnF$_3$ crystallizes in the \emph{Pnma} structure (space group number 62)  which is composed by tilted Mn--F$_6$ octahedra along the three cubic directions and antipolar motions of the Na cations\cite{Daniel1995a}. 
This is the result of the condensation of zone-boundary antiferrodistortive (AFD) unstable modes observed in the cubic reference structure phonon-dispersion curves\cite{acgarciacastro2014}. 
By performing the structural relaxation of the atomic positions and of the cell parameters of NaMnF$_3$ in the $Pnma$ phase, we obtain the lattice parameters $a$ = 5.750 \r{A},  $b$ = 8.007 \r{A} and $c$ = 5.547 \r{A}.
This corresponds to a difference of 0.2$\%$ with respect to the experimental measurements reported by Daniel \emph{et. al.} \cite{Daniel1995a}. 

In Tables \ref{tab:IR-Raman-Pnma} and \ref{tab:pnma-gruneisen} we report the calculated silent, Raman and infra-red (IR) phonon modes as well as the mode Gr\"uneisen parameters ($\gamma_i$) of the relaxed $Pnma$ ground state.
The irreducible representation of the $Pnma$ perovskite phase is $\Gamma = $8A$_u$ $\oplus$ 10B$_{1u}$ $\oplus$ 8B$_{2u}$ $\oplus$ 10B$_{3u}$ $\oplus$ 7A$_{g}$ $\oplus$ 5B$_{1g}$ $\oplus$ 7B$_{2g}$ $\oplus$ 5B$_{3g}$.
Both the Raman (A$_{g}$, B$_{1g}$, B$_{2g}$ and B$_{3g}$ labels) and  IR modes (B$_{1u}$, B$_{2u}$ and B$_{3u}$ labels) are in good agreement with the available experimental reports\cite{Daniel1995a, Lane1971}. 
We note that a very soft polar $B_{2u}$ mode is present with a frequency of 18 cm$^{-1}$.
This indicates that the $Pnma$ phase is close to a ferroelectric phase transition and this mode is responsible for the epitaxial strain induced ferroelectricity of NaMnF$_3$ reported by A. C. Garcia-Castro \emph{et al.}\cite{acgarciacastro2014}
From the Gr\"uneisen parameters reported in Table \ref{tab:pnma-gruneisen}, we observe that two phonon modes are particularly sensitive to a change in volume: the soft B$_{2u}$ mode at 18 cm$^{-1}$ with $\gamma$ = 7.957 and the A$_g$ mode at 79 cm$^{-1}$ with $\gamma$ = 4.844 such as they can be easily detected from Raman experiments. 
Two phonon modes have negative $\gamma$ and correspond to the modes that become soft as the volume decreases, which indicates that probably these two phonon modes are close to a nearest neighbour force instability.

Regarding the magnetic ordering, the total energy differences between the ferromagnetic (FM), $C$- and $A$-type antiferromagnetic (AFM) with respect to $G$-type AFM ordering in the $Pnma$ structure are respectively 79.03, 54.08 and 24.18 meV. 
The lowest energy is thus given by the $G$-type AFM ordering, in agreement with the experimental measurements that establish a $G$-type AFM ordering for the $Pnma$ phase of NaMnF$_3$.\cite{Shane1967} 
All of that confirms that our choice of parameters for the DFT calculations (GGA PBEsol and $U$ = 4.0 eV) of NaMnF$_3$ gives good results when comparing to the experimental measurements.

\begin{table}[htbp!]
\caption{Calculated Raman, Infra-red (IR)  and silent modes frequencies (cm$^{-1}$) of the $Pnma$ phase of NaMnF$_3$ (at zero pressure). 
In brackets are presented the experimental values of the Raman\cite{Daniel1995a} and the IR\cite{Lane1971} modes. }
\centering
\begin{tabularx}{\columnwidth}{X X X X}
\hline
\hline
\multicolumn{4}{c}{Raman modes}         \rule[-1ex]{0pt}{3.5ex} \\
\emph{A$_{g}$} & \emph{B$_{1g}$} & \emph{B$_{2g}$} & \emph{B$_{3g}$} \rule[-1ex]{0pt}{3.5ex} \\
\hline
 79    (88) & 86 (-----)   & 132 (140) & 87 (96) \rule[-1ex]{0pt}{3.5ex} \\
135 (143) & 154 (160) & 145 (155) & 191 (-----) \rule[-1ex]{0pt}{3.5ex} \\
175 (183) & 220 (182) & 191 (201) & 252 (-----) \rule[-1ex]{0pt}{3.5ex} \\
202 (212) & 303 (312) & 216 (226) & 294 (302) \rule[-1ex]{0pt}{3.5ex} \\
242 (250) & 414 (346) & 260 (-----) & 462 (-----) \rule[-1ex]{0pt}{3.5ex} \\
259 (270) &   ---           & 281 (293) &      ---       \rule[-1ex]{0pt}{3.5ex} \\
309 (319) &   ---           & 449 (-----) &      ---       \rule[-1ex]{0pt}{3.5ex} \\
\hline
Silent  & \multicolumn{3}{c}{IR modes }         \rule[-1ex]{0pt}{3.5ex} \\
\emph{A$_{u}$} & \emph{B$_{1u}$} & \emph{B$_{2u}$} & \emph{B$_{3u}$} \rule[-1ex]{0pt}{3.5ex} \\
\hline
59 (-----) & 114 (103) & 18    (-----) & 91 (103) \rule[-1ex]{0pt}{3.5ex} \\
114 (-----) & 125 (132) & 118(120) & 148 (132) \rule[-1ex]{0pt}{3.5ex} \\
124 (-----) & 180 (161) & 180 (181) & 152 (161) \rule[-1ex]{0pt}{3.5ex} \\
133 (-----) & 196 (194) & 205 (-----) & 183 (208) \rule[-1ex]{0pt}{3.5ex} \\
198 (-----) & 223 (228) & 280 (-----) & 222 (220) \rule[-1ex]{0pt}{3.5ex} \\
278 (-----) & 255 (256) & 367 (-----) & 237 (256) \rule[-1ex]{0pt}{3.5ex} \\
363 (-----) & 272 (289) & 387 (-----) & 307 (289) \rule[-1ex]{0pt}{3.5ex} \\
396 (-----) & 312 (-----) &  ---            & 347 (-----) \rule[-1ex]{0pt}{3.5ex} \\
   ---          & 402 (-----) &  ---            & 389 (-----) \rule[-1ex]{0pt}{3.5ex} \\  
\hline
\end{tabularx}
\label{tab:IR-Raman-Pnma}
\end{table}

\begin{table}[htbp!]
\caption{Calculated mode Gr\"uneisen parameters ($\gamma_i$) for the NaMnF$_3$ $Pnma$ phase at $\Gamma$ and around atmospheric pressure.}
\centering
\begin{tabularx}{\columnwidth}{X X X X  c X X X}
\hline
\hline
$A_g$ & $B_{1g}$  & $B_{2g}$ & $B_{3g}$  & $A_u$ & $B_{1u}$ & $B_{2u}$ & $B_{3u}$ \rule[-1ex]{0pt}{3.5ex}\\
\hline
4.844 & 1.425 & 1.226 & 3.017  & -1.018 & 1.674 & 7.957 & 1.752 \rule[-1ex]{0pt}{3.5ex}\\
0.857 & 1.168 & 2.710 & 1.110  & -0.668 & 1.323 & 0.559 & 1.546 \rule[-1ex]{0pt}{3.5ex}\\
2.196 & 2.416 & 2.872 & 1.351  & 0.172 & 0.574 & -0.139 & 0.486 \rule[-1ex]{0pt}{3.5ex}\\
2.273 & 1.714 & 1.401 & 1.833  & 1.692 &  0.707 & 1.663 & 1.717 \rule[-1ex]{0pt}{3.5ex}\\
3.063 & 1.023 & 1.855 & 0.897  & 0.990 &  2.983 & 1.274 & 2.252 \rule[-1ex]{0pt}{3.5ex}\\
1.706 & ---      & 1.627 & ---       & 1.512  & 1.980 & 1.632 & 2.632 \rule[-1ex]{0pt}{3.5ex}\\
1.417 & ---     & 0.919  & ---       & 1.713  & 1.884 & 1.507 & 0.906 \rule[-1ex]{0pt}{3.5ex}\\
---      & ---     & ---       & ---        &  1.452 & 0.992 & ---      & 1.026 \rule[-1ex]{0pt}{3.5ex}\\
---      & ---     & ---       & ---        & ---       & 1.421 & ---      & 1.471 \rule[-1ex]{0pt}{3.5ex}\\
\hline
\end{tabularx}
\label{tab:pnma-gruneisen}
\end{table}

\subsection{Study of NaMnF$_3$ under hydrostatic pressure}

\begin{figure}[htb!]
 \centering
  \subfigure[]
 {
 \includegraphics[width=8.5cm,keepaspectratio=true]{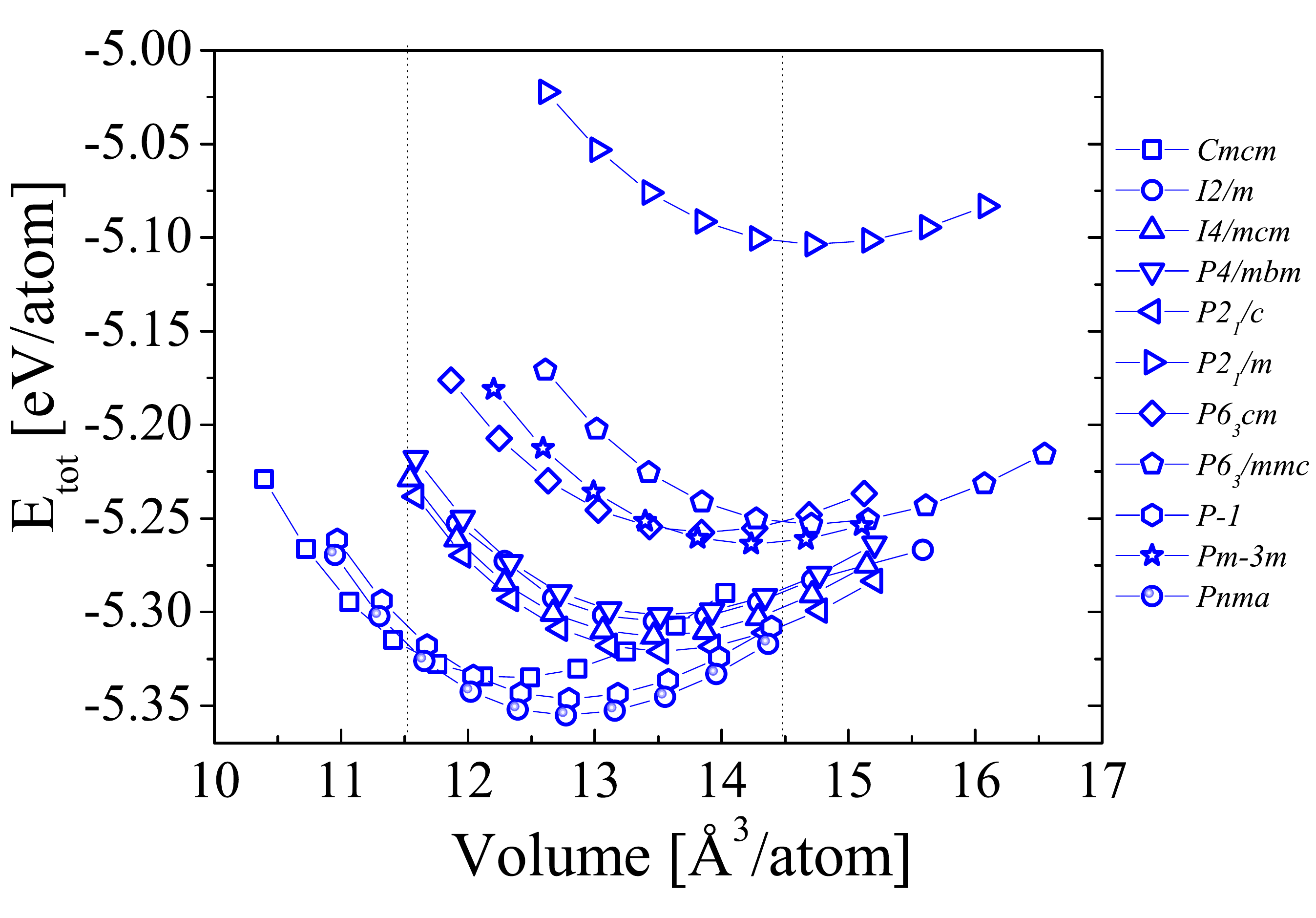}
 }
  \subfigure[]
 {
 \includegraphics[width=8.5cm,keepaspectratio=true]{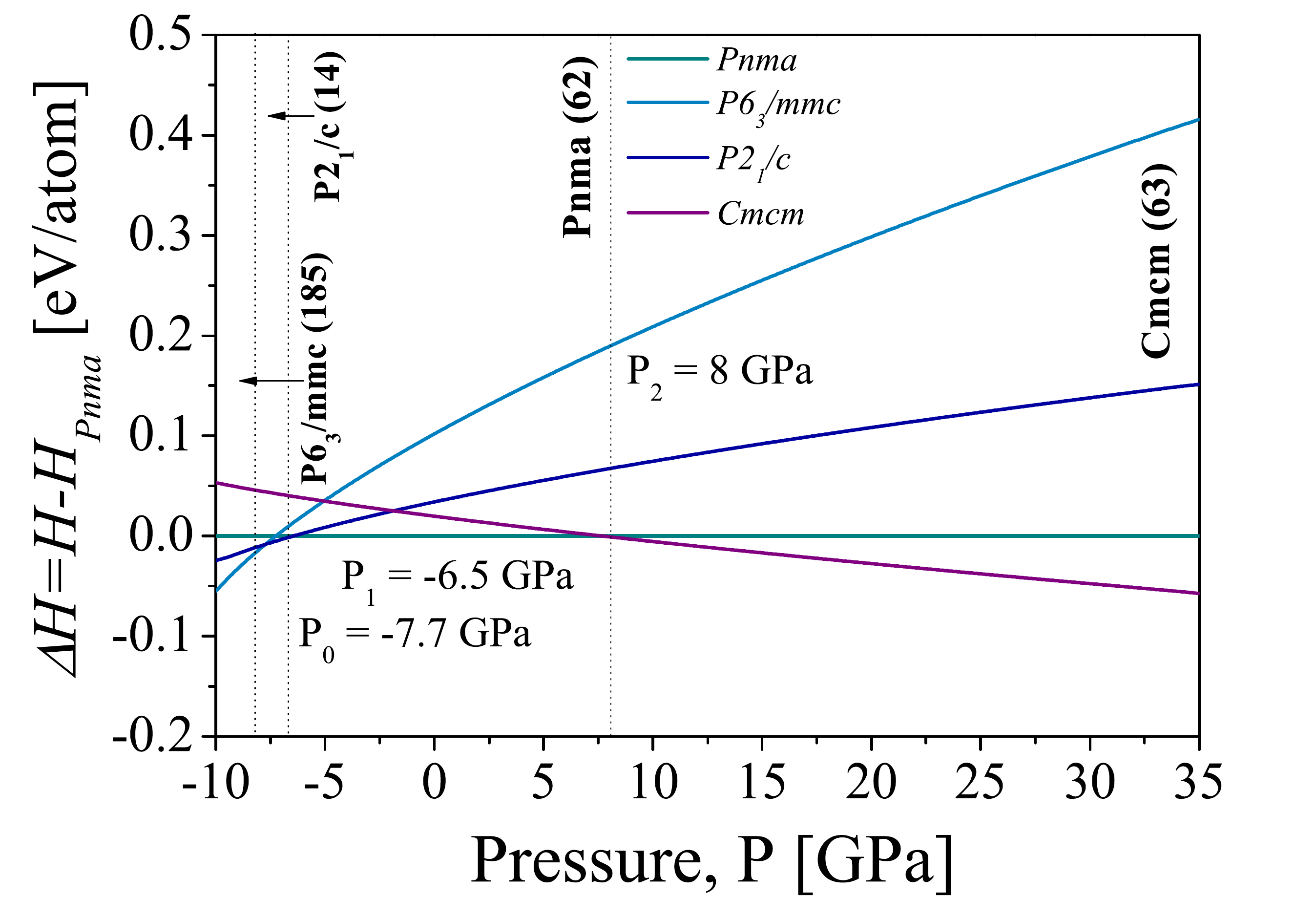}
 }
 \caption{(Color online) (a) Total energy of NaMnF$_3$ as a function of the volume per atom for all different crystal structures. 
 \emph{Cmcm} and \emph{Pnma} structures show the lowest energy values for lower volumes. 
 (b) $\Delta$H/atom as a function of external pressure in NaMnF$_3$ structures. 
 \emph{Pnma} to \emph{Cmcm} structural transition can be seen at the critical pressure of 8 GPa.}
 \label{fig:HvsP}
\end{figure}

In this section we study the pressure phase transitions of NaMnF$_3$. 
In order to identify the enthalpy ground state under hydrostatic pressure, we fitted the total energy versus the volume through the Birch-Murnaghan equation of state (BM-EOS) \cite{Birch1947} for the following common structures found in perovskites: $Cmcm$ ($pPv$ phase), $I2/m$, $I4/mcm$, $P4/mbm$, $P2_1/c$, $P2_1/m$, $P6_3cm$, $P6_3/mmc$, $P\bar{1}$, $Pm\bar{3}m$, and $Pnma$. 
For all of them, the $G$-type AFM magnetic ordering was considered in our calculations. 
In Fig.\ref{fig:HvsP}a we report the total energy values per atom as a function of the volume change with respect to the equilibrium volume.
As we would expect, the \emph{Pnma} and \emph{Cmcm}-$pPv$ structures exhibit the lowest energy values at low volumes. 
In order to identify the lowest energy phase, we report the enthalpy (H), as obtained from the equation of state, versus the pressure  in Fig.\ref{fig:HvsP}b. 
Going from the lowest to the highest pressure, we found the following low enthalpy phases.
Between -10 GPa and -8.9 GPa the hexagonal $P6_3/mmc$  phase is the most stable structure.
Between  -8.9 GPa and -6.5 GPa we found the $P2_1/c$ structure to be the most stable.
Then between -6.5 GPa and 8.0 GPa the orthorhombic \emph{Pnma} structure is the ground state and thus passing through the atmospheric pressure as expected from the experimental results. 
At higher pressures (above 8 GPa, up to 35 GPa) the orthorhombic \emph{Cmcm} $pPv$ structure is the lowest energy phase. 
Then, we found that NaMnF$_3$ presents a phase transition at relatively low pressure (8 GPa) from the \emph{Pnma} perovskite phase to the \emph{Cmcm} $pPv$ phase.

\begin{figure}[htb!]
 \centering
 \subfigure[]
 {
 \includegraphics[width=7.0cm,keepaspectratio=true]{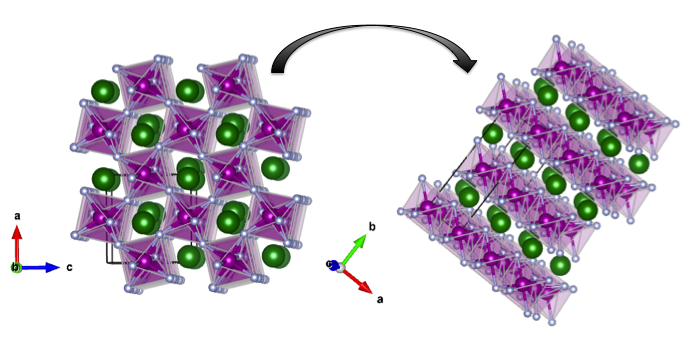}
 }
  \subfigure[]
 {
 \includegraphics[width=6.8cm,keepaspectratio=true]{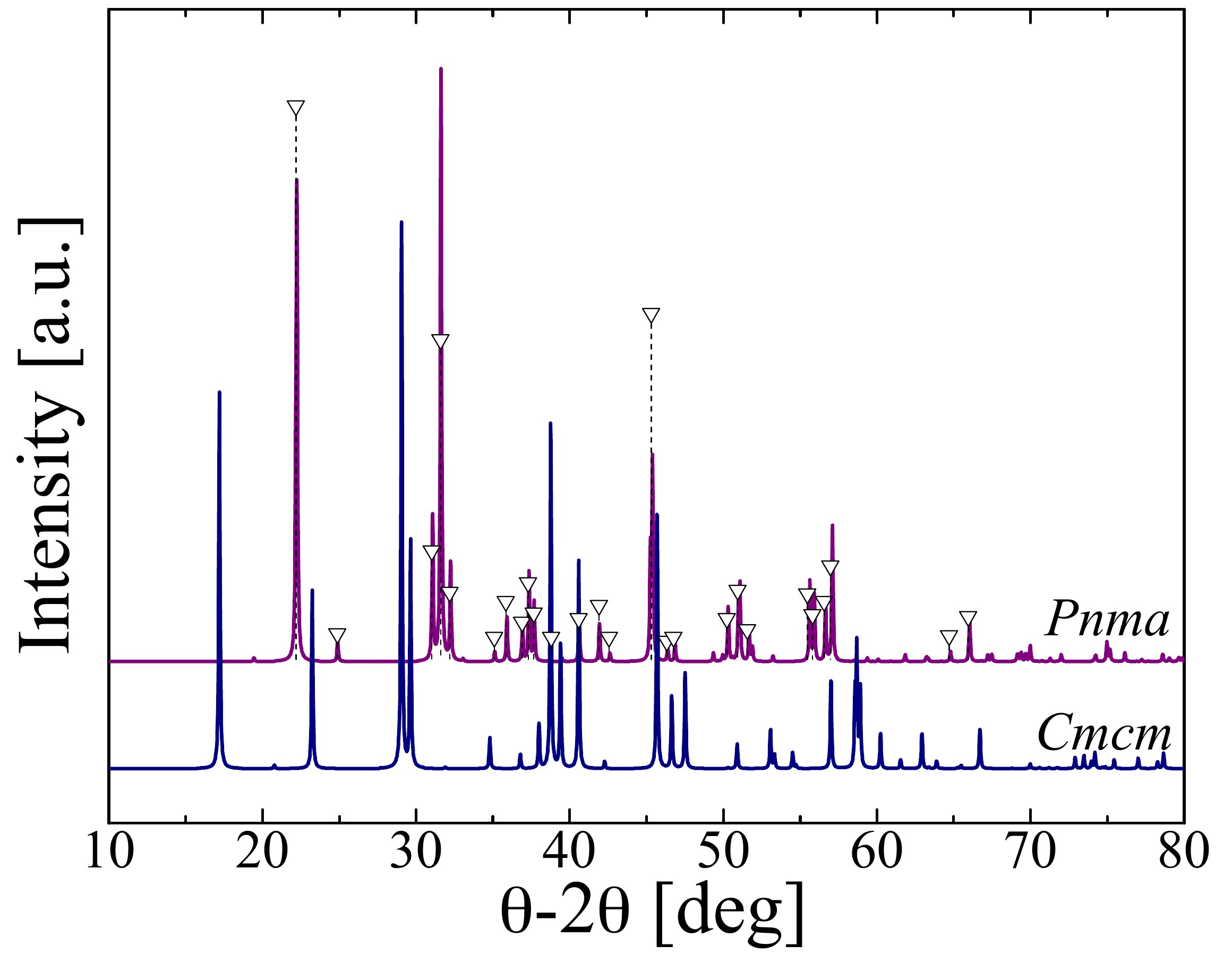}
 }
  \subfigure[]
 {
 \includegraphics[width=7.8cm,keepaspectratio=true]{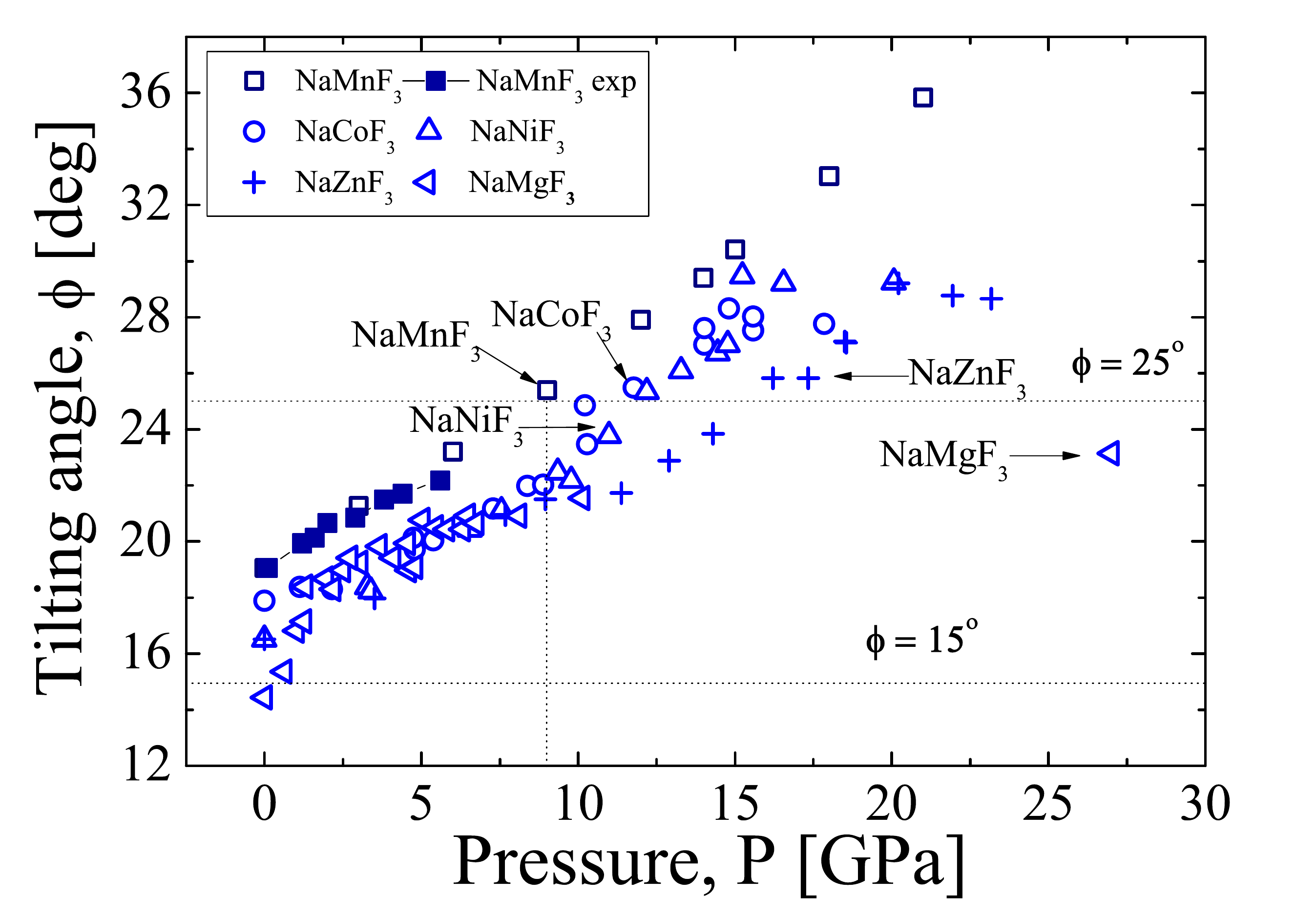}
 }
 \caption{(Color online) (a) Schematic view of the phase transition from \emph{Pnma} to \emph{Cmcm}. 
 Na, Mn and F atoms are represented in green, violet and grey colors respectively. 
 The breaking of bondings for the transition from corner-shared to edge-shared octahedra can be appreciated. 
 (b) Simulated XRD patterns for $Pnma$ and $Cmcm$ structures. 
 The room temperature $Pnma$ phase is in good aggreament with the experimental reports (open triangles).\cite{Akaogi2013, Daniel1995a, Lane1971} 
 (c) Octahedra tilt angle as a function of pressure in Na$M$F$_3$. 
 For $M$ = Zn, Ni, Co and Mg experimental results are taken from Yusa \emph{et. al.} \cite{Yusa2012}. 
 All of these compounds present the transition from the \emph{Pnma} to the \emph{Cmcm} structures at a tilting angle close to 25$^{\circ}$. 
 Experimental results for NaMnF$_3$ at low pressure were taken from Katrusiak \emph{et. al.}\cite{Katrusiak1992} 
 Full agreement between experimental (filled squares) and our theoretical calculations (empty squares) can be appreciated.
 }
\label{fig:tilt-ABF3}
\end{figure}

In Fig.\ref{fig:tilt-ABF3}a we represent a schematic view of the $Pnma$ and the $pPv$ phases. 
It is clear that the structural transition between the $Pnma$ and the $pPv$ phase is reconstructive since a breaking of Mn--F--Mn bonds is needed to connect the two structures.
This corresponds to a transition between corner-shared octahedra in the $Pnma$ phase to edge-shared octahedra in the $pPv$ phase.
The $pPv$ phase can be viewed as layers of edge-shared octahedra separated by layers of Na cations along the $y$-direction.
In Fig.\ref{fig:tilt-ABF3}b we report the calculated X-ray diffraction patterns (XRD) for the $Pnma$ and the post-perovskite $Cmcm$ phases with a Cu-K$\alpha$ =  1.5418 \r{A} X-ray wavelength. 
The XRD pattern of the room temperature $Pnma$ is in good agreement with the experimental reports shown as a triangles in the Fig.\ref{fig:tilt-ABF3}b.\cite{Akaogi2013, Daniel1995a, Lane1971}
No experimental data exists for the $pPv$ phase of NaMnF$_3$ but the calculated XRD pattern fit the $pPv$ symmetry found in other compounds such as MgGeO$_3$.\cite{Hirose2005}

An interesting parameter to be analysed in the $Pnma$ to $pPv$ phase transition is the dependence of the octahedral tilting angle of the \emph{Pnma} structure as a function of pressure.  
O'Keeffe \emph{et. al.} reported that the transition from the $Pnma$ to the $pPv$ structure takes place when the octahedra tilting angles reach 25$^{\circ}$.\cite{O'keeffe1979}
This condition is reached when the atmospheric pressure tilting angle of the $Pnma$ phase is above 15$^{\circ}$.\cite{O'keeffe1979, Tsuchiya2004, Tateno201054} 
For this purpose,  we computed the pressure dependence of the octahedra titling angle $\phi$ of the \emph{Pnma} structure following to the relationship deduced by O'Keeffe \emph{et. al.} \cite{O'keeffe1979} such as $\phi$ = $cos^{-1}(\sqrt{2}c^2/ab)$ with $a$, $b$ and $c$ are the cell parameters.
We report our results as a function of pressure for NaMnF$_3$ and the experimental data for Na$M$F$_3$ compounds with $M$ = Zn, Ni, Co, Mg,  and Mn in Fig.\ref{fig:tilt-ABF3}c. 
The experimental results for $M$ = Zn, Ni, Co and Mg are taken from Yusa \emph{et. al.} \cite{Yusa2012}. 
Experimental data for lattice parameters used in the calculation of $\phi$  were extracted from the report of Katrusiak \emph{et. al.}\cite{Katrusiak1992}
As it can be appreciated, all the experimental measurements follow the tilting criterion proposed by O'Keeffe, \emph{i.e.} a $pPv$ phase transition at $\phi=$25$^\circ$.
The experimental results for NaMnF$_3$ (filled squares) are in good agreement with our calculations (open squares) up to 5.6 GPa (maximum pressure explored experimentally). 
Our calculations predict that the tilt angle reaches 25$^\circ$ at 8 GPa, where the transition should occur according to the  O'Keeffe rule which is indeed in agreement with our BM-EOS prediction of 8 GPa as critical pressure.

Our theoretical findings for the transition between the $Pnma$ and the $Cmcm$ $pPv$ phase do not follow the recent experimental report of Akaogi \emph{et. al.} \cite{Akaogi2013}. 
In their pressure measurements at high temperature (T = 1273 K),  Akaogi \emph{et. al.}  do not observe a transition from the $Pnma$ to the $pPv$ structure up to 24 GPa but found, instead, a MnF$_2$ + Na$_3$Mn$_2$F$_7$ phase decomposition around 8 GPa.
We note that Katrusiak \emph{et al.} also report a transition from the $Pnma$ to the cubic phase $Pm\bar{3}m$ at T$_c$ = 970 K and ambient pressure\cite{Katrusiak1992} for this compound which is not observed by Akaogi \emph{et. al.} at very high temperature (higher than T$_c$ = 970K). 
We suggest that at the very high temperature where Akaogi \emph{et. al.} performed their measurements, it is indeed possible that a phase decomposition appears such as they do not observe the $pPv$ transition. 
We also remark that at such high temperature, the O'Keeffe rule is not fulfilled since the crystal is cubic or with titling angle smaller than the required critical value of 15$^\circ$ to observe the $Pnma$ to $pPv$ phase transition under pressure such as our calculation are still valid at low temperature.
Following the experimental results and in order to understand the decomposition phases, we performed calculations for Na$_3$Mn$_2$F$_7$ and MnF$_2$  compounds. 
The first stoichiometry is the one of the Ruddlesen-Popper (RP) phases\cite{rp2000} with \emph{A$_{n-1}$A'$_2$B$_n$X$_{3n+1}$} formula, which in our particular case corresponds to $n$ = 2 and $A'$ = $A$ = Na, $B$ = Mn and $X$ = F. 
To our knowledge, there is no previous report of the sodium-manganese based compound in the RP phase. 
Therefore, all reported phases for similar compounds (oxides\cite{rp2000} and fluorides\cite{BABEL198577}) were tested and the energy minimum per structure was analysed. 
Nevertheless, none of our simulated XRD patterns of the considered RP phases match the one reported by Akaogi \emph{et. al.}\cite{Akaogi2013}, while for the MnF$_2$ compound, we obtain a good agreement with the experiments.\cite{Akaogi2013}  
We are thus confident about our predicted phase transition from the perovskite toward the $pPv$ phase of NaMnF$_3$ at relatively low pressure (8 GPa) and more experimental measurements would be required in order to clarify the pressure phase transition that takes place in NaMnF$_3$ at lower (room) temperature.

\begin{table}[htbp!]
\caption{Computed Raman, Infra-red and silent modes of NaMnF$_3$ in post-perovskite \emph{Cmcm} phase as well as the mode Gr\"uneisen parameters ($\gamma_i$) at $\Gamma$ around 12 GPa.}
\centering
\begin{tabularx}{\columnwidth}{X X X X  | X X X X}
\hline
\hline
\multicolumn{8}{c}{Raman, IR and silent modes [cm$^{-1}$]}         \rule[-1ex]{0pt}{3.5ex} \\
\emph{A$_{g}$} & \emph{B$_{1g}$} & \emph{B$_{2g}$} & \emph{B$_{3g}$} & \emph{A$_{u}$} & \emph{B$_{1u}$} & \emph{B$_{2u}$} & \emph{B$_{3u}$} \rule[-1ex]{0pt}{3.5ex} \\
\hline
194 & 118 & 284 & 138  & 62 & 90   & 196 & 125 \rule[-1ex]{0pt}{3.5ex}\\
239 & 240 & --- & 228  & 335 & 168 & 223 & 248 \rule[-1ex]{0pt}{3.5ex}\\
310 & 298 & --- & 358  & ---   & 192 &  294 &  343 \rule[-1ex]{0pt}{3.5ex}\\
402 & ---   & --- & 512  & ---   & 356 & 374 & --- \rule[-1ex]{0pt}{3.5ex}\\
---   & ---   & --- & ---     & ---   & 509 & 416 & --- \rule[-1ex]{0pt}{3.5ex}\\
\hline
\multicolumn{8}{c}{Mode Gr\"uneisen parameters ($\gamma_i$)}         \rule[-1ex]{0pt}{3.5ex} \\
1.619 & 0.702 & 1.891 & 1.674  & 1.699 & -0.137 & -0.204 & 0.074 \rule[-1ex]{0pt}{3.5ex}\\
0.757 & 1.623 & ---     & 1.975   & 1.293 & -0.204 & 0.449 & 2.011 \rule[-1ex]{0pt}{3.5ex}\\
1.265 & 2.107 & ---     & 1.301   & ---      & 2.558 & 1.427 & 1.384 \rule[-1ex]{0pt}{3.5ex}\\
1.050 & ---      & ---     & 1.171   & ---      & 1.265 & 0.972 & --- \rule[-1ex]{0pt}{3.5ex}\\
---      & ---      & ---     & ---         & ---      & 1.543 & 1.156 & --- \rule[-1ex]{0pt}{3.5ex}\\
\hline
\end{tabularx}
\label{tab:IR-Raman-Cmcm}
\end{table}

\begin{figure}[htb!]
 \centering
 \includegraphics[width=8.5cm,keepaspectratio=true]{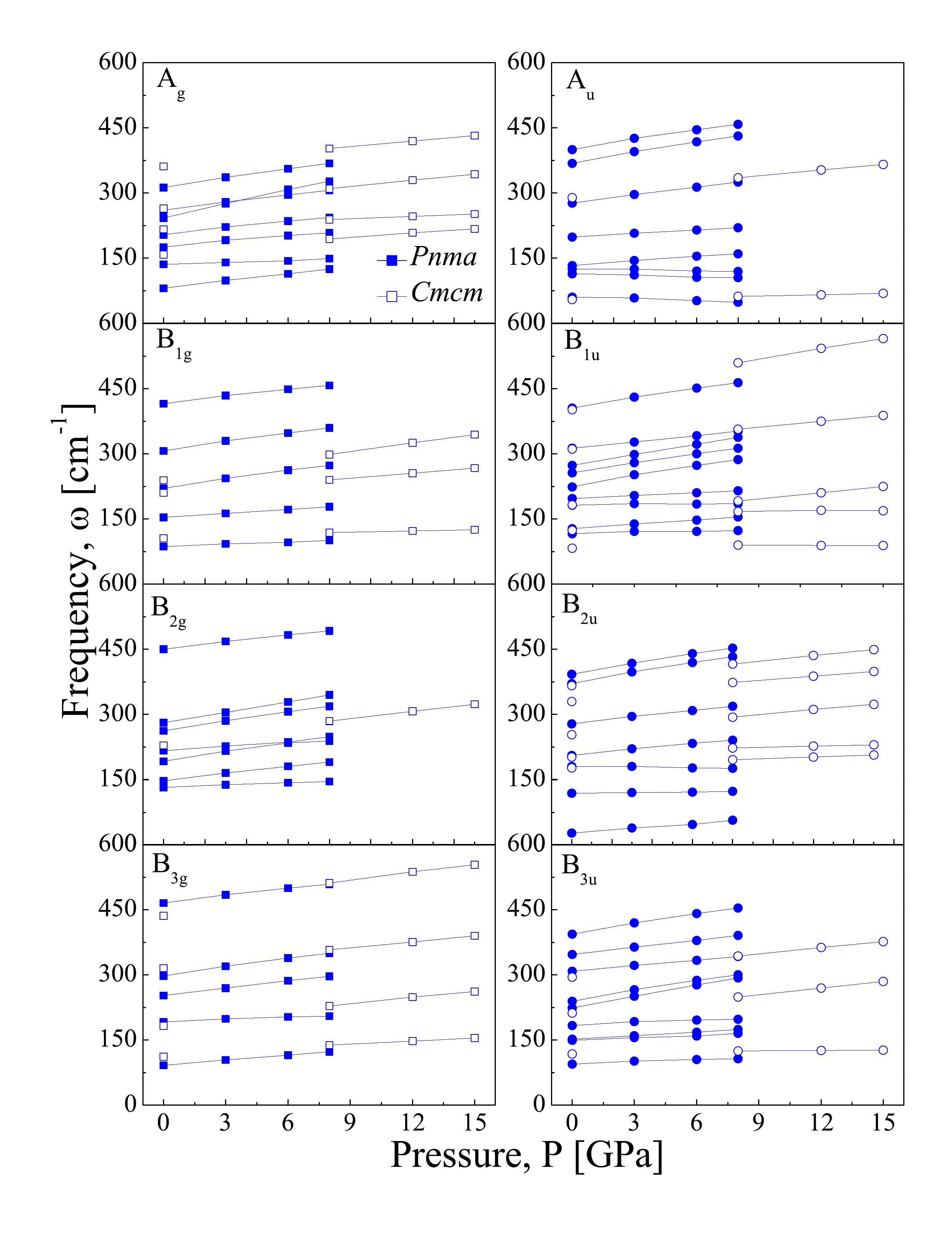}
 \caption{(Color online) Vibrational behavior as a function of isotropic pressure for $Pnma$ and $Cmcm$ $pPv$. The open and filled symbols are the Raman and the Infra-red modes respectively. }
 \label{fig:w-vs-p}
\end{figure}

We now focus on the vibrational characterization of the $pPv$ phase at 8 GPa. 
In Table \ref{tab:IR-Raman-Cmcm} we report the computed phonon mode frequencies and the mode Gr\"unisen parameter ($\gamma_i$) in the \emph{Cmcm} $pPv$ structure with the lattice parameters relaxed at 8 GPa  (also reported in Table \ref{tab:Cmcm-all}).
The irreducible representation at the $\Gamma$ zone center point is: 2A$_u$ $\oplus$ 6B$_{1u}$ $\oplus$ 6B$_{2u}$ $\oplus$ 4B$_{3u}$ $\oplus$ 4A$_g$ $\oplus$ 3B$_{1g}$ $\oplus$ B$_{2g}$ $\oplus$ 4B$_{3g}$. 
Two silent modes ($A_u$ label), sixteen IR active modes ($B_{1u}$, $B_{2u}$ and $B_{3u}$) and twelve Raman active modes ($B_{1g}$, $B_{2g}$ and $B_{3g}$) are thus expected. 
As we can see in Table \ref{tab:IR-Raman-Cmcm}, all the calculated modes at the $\Gamma$ point of the BZ are positive, indicating a vibrational local stability of the $pPv$ phase at 8GPa. 
The lowest frequency mode is a silent mode $A_u$ at 62 cm$^{-1}$ and the lowest polar mode is a $B_{1u}$ mode with a frequency of 90 cm$^{-1}$.
The obtained Gr\'uneisen parameters are smaller than the one obtained in the $Pnma$ ground state structure, which can be explained from an interatomic force stiffness larger in the $pPv$ phase than in the $Pnma$ phase.
Even though there are some phonon modes with negative parameter, they are not large enough to have an effect on the thermal expansion.
We also report in Fig.\ref{fig:w-vs-p} the evolution of the Raman and IR modes as a function of pressure from 0 GPa to 15 GPa.
After the transition, the system goes to a reduced crystallographic unit cell such as the number of modes is two times smaller in the $Cmcm$ $Ppv$ phase than it is in the $Pnma$ phase.
We note that whatever the $Pnma$ or the $pPv$ structure, all the phonon mode frequencies harden when increasing the pressure.
This is also valid for the soft $B_{2u}$ polar mode of the $Pnma$ structure that hardens as the pressure increases and thus in agreement with the usual behaviour of perovskites in the presence of the AFD distortions, which are inimical to the ferroelectric distortions and are enhanced with pressure.

\subsection{Magnetic properties of the $pPv$ phase}
Before  performing the non-collinear magnetic calculations, it is useful to make a preliminary symmetry analysis in order to determine the magnetic orderings that are allowed in the $pPv$ structure.  
The $Cmcm$ space group belongs to the \emph{D$_{2h}$} point group~\cite{Arroyo1, Arroyo2} and the Mn atoms are placed at the $4c$ Wyckoff position with coordinates (0,0,0) for Mn$_1$, (0,0,$\frac{1}{2}$) for Mn$_2$, ($\frac{1}{2}$,$\frac{1}{2}$,0) for Mn$_3$ and ($\frac{1}{2}$,$\frac{1}{2}$,$\frac{1}{2}$) for Mn$_4$.
From these 4 magnetic cations, we can define 4 collinear magnetic orderings depending on the sign of the magnetic moments of (Mn$_1$, Mn$_2$, Mn$_3$, Mn$_4$): \emph{F} ($+$, $+$, $+$, $+$) for the FM ordering, \emph{A} ($+$, $+$, $-$, $-$) for the $A$-type AFM ordering, \emph{C} ($+$, $-$, $+$, $-$)  for the $C$-type AFM ordering and \emph{G} ($+$, $-$, $-$, $+$) for the $G$-type AFM ordering, where $+$ and $-$ correspond to spin up and down respectively. 
Of course, the ferrimagnetic ordering ($+$, $+$, $+$, $-$) could be considered but it is unlikely to appear in real conditions because of the symmetry of the system and the superexchange interaction.
For the non-collinearity we have to consider the three Cartesian directions $x$, $y$ and $z$ for each magnetic moment which gives 4$\times$3 possible states: $F_x$, $F_y$, $F_z$, $A_x$, $A_y$, $A_z$, $C_x$, $C_y$, $C_z$, $G_x$, $G_y$ and $G_z$. 
Then, taking into account the fact that the spins are axial vectors, one has to apply the symmetry operations of the \emph{D$_{2h}$} point group on each of the non-collinear magnetic orders defined above and see how they transform.
The result of this approach applied on the $pPv$ phase of NaMnF$_3$ is summarised in Table \ref{tab:ord-sym}.

\begin{table}[htbp!]
\caption{Magnetic ordering allowed in the $Cmcm$ structure according to the \emph{D$_{2h}$} point group.\cite{Arroyo1, Arroyo2}
The transformation of each magnetic ordering under each symmetry operation is labeled as +1 and -1, which indicates when the ordering is invariant under the related transformation ($i.e.$ +1) or when it is reversed under the application of the symmetry operation ($i.e.$ -1).}
\centering
\begin{tabularx}{\columnwidth}{X  c  c  c  c  c  c  c  c  c}
\hline
\hline
Magnetic Ordering & \emph{I} &  \emph{2$_z$} &  \emph{2$_y$} &  \emph{2$_x$} &  \emph{-1} &  \emph{m$_z$} &  \emph{m$_y$} &  \emph{m$_x$} &    \rule[-1ex]{0pt}{3.5ex} \\
\hline
\emph{F$_x$}, \emph{A$_x$} &  1  &   -1  &  -1  &  1  &  1  &  -1  &  -1  &  1   &  \emph{B$_{3g}$} \rule[-1ex]{0pt}{3.5ex}\\
\emph{F$_y$}, \emph{A$_y$}, \emph{C$_z$}, \emph{G$_z$} &  1  &   -1  &  1  &  -1  &  1  &  -1  &  1  &  -1   &  \emph{B$_{2g}$} \rule[-1ex]{0pt}{3.5ex}\\
\emph{F$_z$}, \emph{A$_z$}, \emph{C$_y$}, \emph{G$_y$} &  1  &   1  &  -1  &  -1  &  1  &  1  &  -1  &  -1   &  \emph{B$_{1g}$} \rule[-1ex]{0pt}{3.5ex}\\
\emph{C$_x$}, \emph{G$_x$}, &  1  &   1  &  1  &  1  &  1  &  1  &  1  &  1   &  \emph{A$_g$} \rule[-1ex]{0pt}{3.5ex}\\
\hline
\end{tabularx}
\label{tab:ord-sym}
\end{table}

As we can see from Table \ref{tab:ord-sym}, the $F_x$ and $A_x$ orders belong to the same character.
This means that in an energy expansion with respect to these two order parameters the second order term $F_x\cdot A_x$ is invariant by symmetry.
This implies that if one of the two order parameters condenses in the structure, the second one can develop as well but as a secondary or a slave order (pseudo-proper).
This secondary order parameter usually appears as a spin canting in the real system and its intensity will depend on the amplitude of the $F_x\cdot A_x$ coupling.
Going through the other magnetic orders, in Table \ref{tab:ord-sym} we also observe that $F_y$, $A_y$, $C_z$ and $G_z$ belong to the same $B_{2g}$ character, $F_z$, $A_z$, $C_y$ and $G_y$ belong to the $B_{1g}$ character and $C_x$ and $G_x$ belong to the $A_g$ character.
Interestingly, whatever the main magnetic order we will observe in the $pPv$ phase of NaMnF$_3,$ a canting of the spins is always possible.
Additionally, we also remark that three of the four possible characters to which belong all the possible magnetic orders, three of them allow for ferromagnetism and so have the potential to exhibit weak FM.

\begin{table}[htbp!]
\caption{Comparison of the distances and the bond angles along the Mn--F--Mn bonds between the \emph{Pnma} and the \emph{Cmcm} structures of NaMnF$_3$.}
\centering
\begin{tabularx}{\columnwidth}{c  X  c | X  c}
\hline
\hline
\emph{Distance and Angle} &   \multicolumn{2}{c}{$Pnma$}      &   \multicolumn{2}{c}{$Cmcm$} \rule[-1ex]{0pt}{3.5ex} \\
\hline
                                                    & [101]    &   3.9951      &      [100]   &   2.9903      \rule[-1ex]{0pt}{3.5ex}\\
\emph{d$_{Mn-Mn}$} [\r{A}]        & [10-1]    &   3.9951      &      [001]   &   3.6644      \rule[-1ex]{0pt}{3.5ex}\\
                                                    & [010]    &   4.0036      &      [010]   &   4.9731      \rule[-1ex]{0pt}{3.5ex}\\
\hline
                                                    & [101]    &   142.58      &      [100]   &   92.46      \rule[-1ex]{0pt}{3.5ex}\\
\emph{$\gamma$ $_{Mn-F-Mn}$} [deg]     & [10-1]          &   142.58      &      [001]   &     134.55    \rule[-1ex]{0pt}{3.5ex}\\
                                                    & [010]    &   140.14      &      [010]   &   ---      \rule[-1ex]{0pt}{3.5ex}\\
\hline
\end{tabularx}
\label{tab:dis-and-angles}
\end{table}

\begin{figure}[htb!]
 \centering
 \includegraphics[width=8.2cm,keepaspectratio=true]{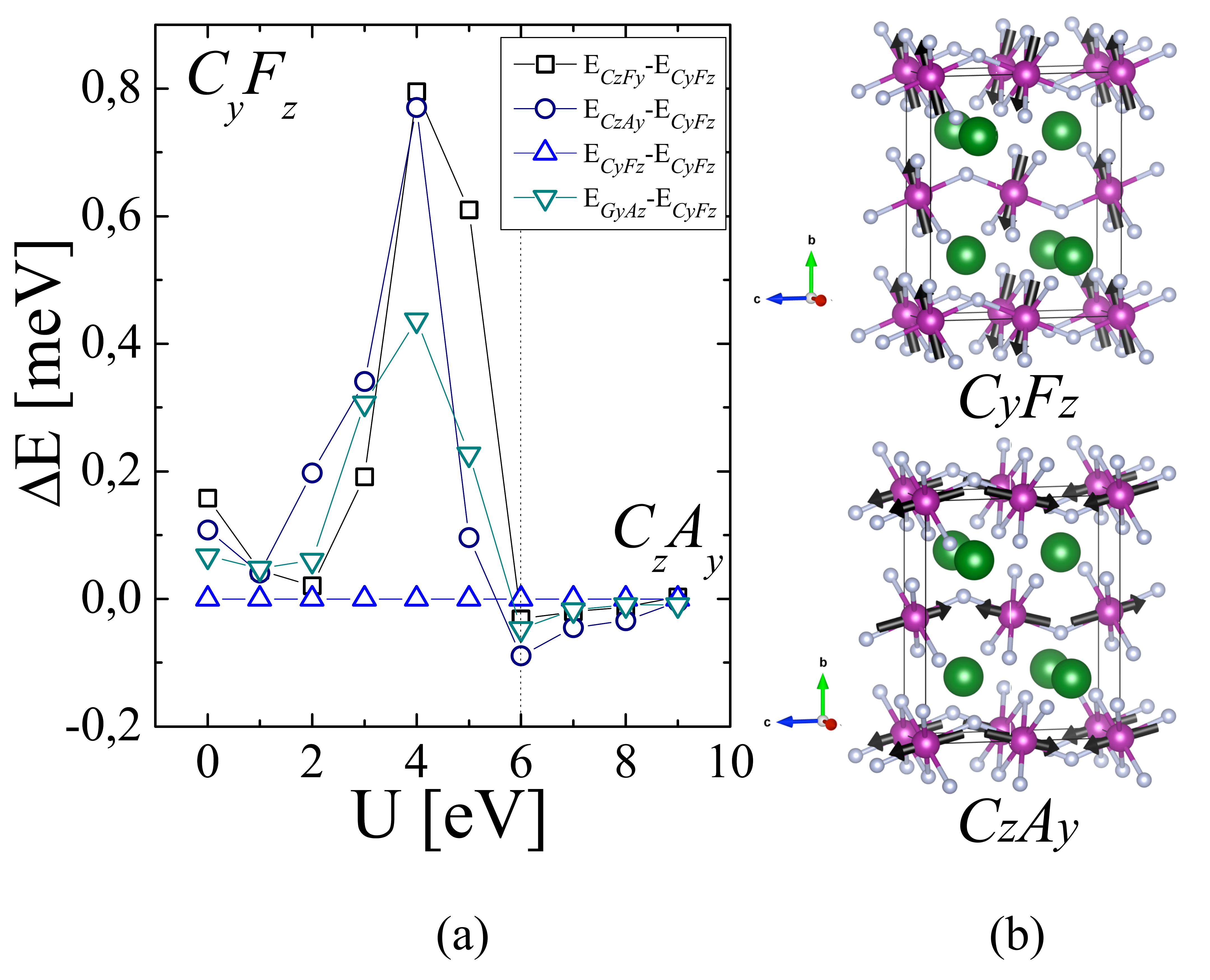}
 \caption{(Color online) (a) Energy differences between the different magnetic orderings of the $pPv$ phase of NaMnF$_3$ versus the $U$ parameter. 
 (b)  Sketch of the \emph{F$_z$C$_y$} and the \emph{A$_y$C$_z$}  magnetic orderings in the $pPv$ phase of NaMnF$_3$. 
 For $U$ < 6.0 eV the system exhibits a AFM+FM magnetic state with a high ferromagnetic canting. 
 For $U$ > 6.0 eV, the system transit toward a canted AFM state with no weak FM.}
 \label{fig:magnetic-transition}
\end{figure}

As discussed above, NaMnF$_3$ compound in the \emph{Pnma}  phase has a $G$-type AFM ordering.\cite{Katrusiak1992, Shane1967, Daniel1995a} 
In Table \ref{tab:dis-and-angles} we report the computed Mn--F--Mn distances and angles of the $Pnma$ phase at room pressure and  of the $pPv$ phase at 8 GPa.  
We note that in the \emph{Pnma} structure, the Mn--Mn bonding in the three directions have similar Mn--F--Mn angles (about 140$^\circ$) and distances (about 4 \AA) while their are much more anisotropic in the $pPv$ phase (angles of 92$^\circ$ and 134$^\circ$ and one non-existent angle due to the octahedral edge sharing of the structure with distances of 2.99 \AA\ and 3.66 \AA\ in the $x-z$ plane and 4.97 \AA\ in the $y$ direction).
Having angles closer to 180$^\circ$ in the $Pnma$ explains the $G$-type AFM ordering since with such bond angle the superexchange interaction between the Mn-$d^5$ spins favours an AFM alignment.\cite{Kanamori1957, Kanamori1957a, Goodenough1958}
In the $pPv$ case, we can see that there is no direct bonding between the Mn atoms along the $y$ direction such as we can expect a weak spin coordination between the spins along the $y$ direction and one would expect a quasi-2D magnetic behaviour in the $pPv$ structure. 
In the plane perpendicular to the $y$ direction we have two types of bonding giving rise to an angle close to 180$^\circ$ along the $z$ direction and to an angle close to 90$^\circ$ along the $x$ direction.
On the top of that, we also note that the 90$^\circ$ bonding along the $x$ direction is given through two bonds since it is the direction where the octahedra share their edge.
The superexchange interaction should thus drives strong FM ordering along the $x$ axis and AFM ordering along the $z$ axis and a weak interaction along the $y$ axis.
Interestingly, the magnetic ordering that fulfils these superexchange rules is the $C$-type AFM and it is indeed the lowest energy collinear magnetic ordering that we found in our calculations (see Fig.\ref{fig:magnetic-transition}b).
Therefore, we can only consider the set of non-collinear calculations containing the $C$-type AFM: ($C_z$, $F_y$, $G_z$, $A_y$), ($C_y$, $F_z$, $A_z$, $G_y$) and ($C_x$, $G_z$) as predicted by group theory. 
We can also expect the $C_x$ case to be unlikely observed since it forces the spins to be antiferromagnetically aligned along the $x$ direction while a strong FM interaction is expected from the double 90$^\circ$ bond angle connecting the Mn atoms in this direction.

\begin{figure}[htb!]
 \centering
 \includegraphics[width=8.7cm,keepaspectratio=true]{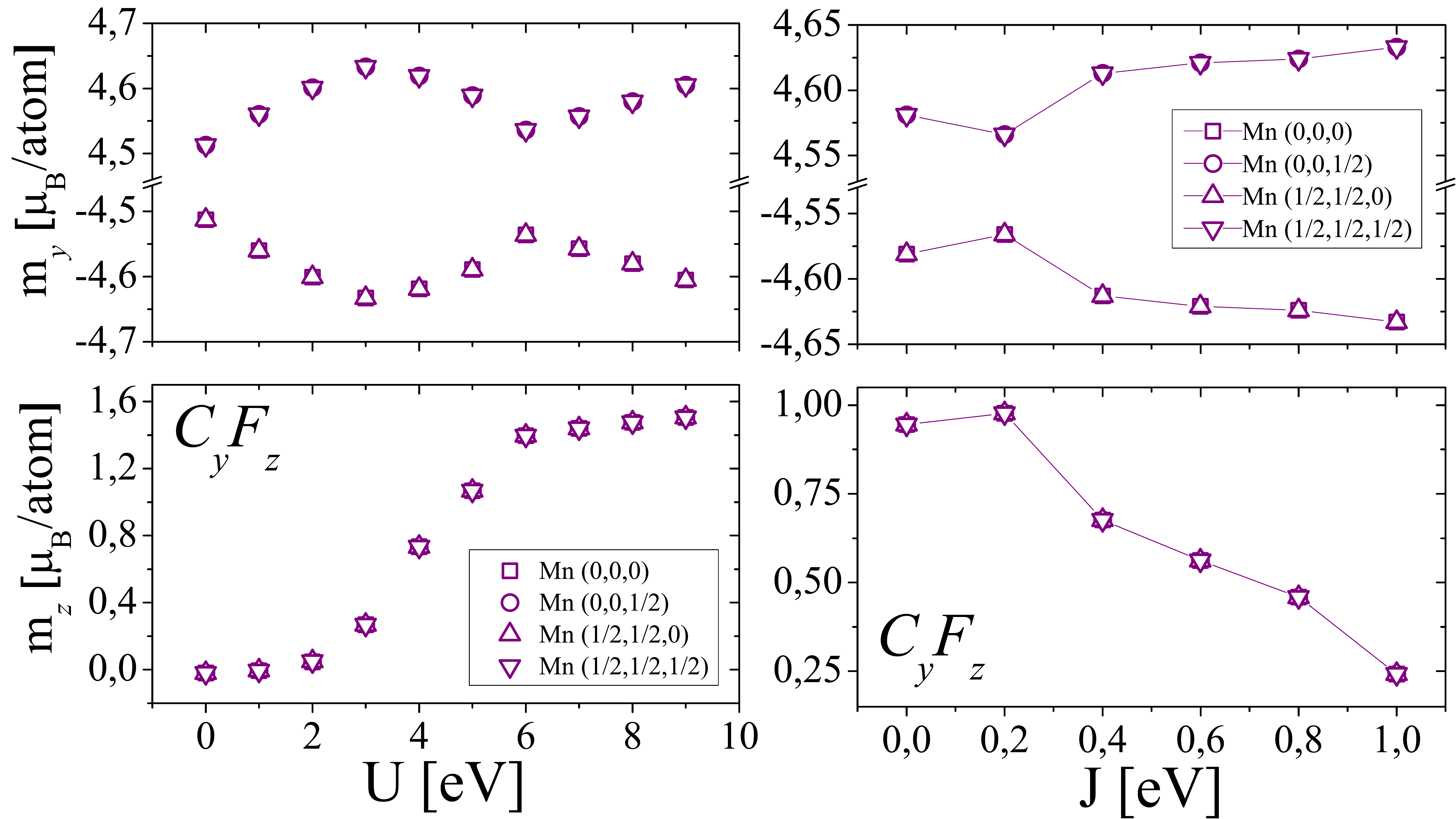}
 \caption{(Color online) Magnetic canting dependence as a function of $U$ and $J$ parameters for NaMnF$_3$. 
 Canting angle ($\varphi$) varies from 0.11$^{\circ}$ to 17.94$^{\circ}$ when $U$ goes from 0.0 to 9.0 eV. 
 The canting angle goes from 12.07$^{\circ}$ to 3.01$^{\circ}$ when $J$ varies from 0.0 to 1.0 eV.}
 \label{fig:mag-vs-U}
\end{figure}

Performing the spin relaxations within the non-collinear regime and within the $C$-type AFM ordering, we found that the lowest energy is obtained for the $C_y$ orientation with a large $F_z$ component along the $z$ direction (no canting along the $x$ axis is observed).
This magnetic ground state is however dependent on the $U$ and $J$ values of the GGA+$U$ exchange correlation functional.
In Fig.\ref{fig:magnetic-transition}a we report the energy difference between the $C_yF_z$ ordering and the other lowest energy cases $C_zF_y$, $C_zA_y$, and $G_yA_z$ versus the $U$ value (fixing $J = 0$).
Here we remark that the $C_yF_z$, $C_zF_y$, $C_zA_y$, and $G_yA_z$ states are very close in energy such as we can expect an easy magnetic phase transitions with small perturbations such as an applied magnetic field.
Such a transition under a magnetic field has been observed experimentally in the $pPv$ phase of NaNiF$_3$ where a field parallel to the $y$ axis induces a transition to a AFM+weak-FM ground state.\cite{Shirako2012}
We can see from Fig.\ref{fig:magnetic-transition}a that the $C_yF_z$ corresponds to the lowest energy state from $U$ = 0.0 eV to $U$ = 5.0 eV while from $U$ = 6.0 eV to $U$ = 9.0 eV the $C_zA_y$ is observed to be the most stable.
This shows that a magnetic spin flop transition appears in the $pPv$ phase of NaMnF$_3$ as function of the electronic Coulomb direct exchange $U$. 
In Fig.\ref{fig:mag-vs-U} we report the evolution of the spin canting with respect to $U$.
The amplitude of the $F_z$ canting goes from 0.01 to 1.6 $\mu_B$/atom when varying $U$ from 0.0 to 9.0 eV.
We note that the $F_z$ component strongly increase from $U$ = 0.0 to 6.0 eV and reach a saturation value beyond $U$=6.0 eV.
Unfortunately, no experimental data of the non-collinearity of NaMnF$_3$ in its $pPv$ phase has been reported in the literature such as it is not possible to validate which $U$ value would be the best to reproduce the spin canting amplitude.
In previous studies with $d^5$ half filled orbitals the range of $U$ values that have been found to treat correctly the property of the systems is around $U$ = 4.0 eV.
It is important to note that the pressure calculations performed with $U$ = 4.0 eV in the $Pnma$ and $pPv$ phase give the best agreement with experiments such as we will use $U$=4.0 eV in the rest of our analysis.

Beside the $U$ parameter, for non-collinear magnetism within the DFT+$U$ approach it is also necessary to check the dependence with respect to the $J$ parameter.\cite{bousquet2010b}
In Fig.\ref{fig:mag-vs-U} we present the evolution of the spin canting versus the $J$ parameter in the $C_yF_z$ ground state of the $pPv$ phase of NaMnF$_3$ with $U$= 4.0 eV.
As we can see, the FM component $F_z$ is strongly affected by $J$, going from 0.95 $\mu_B$ to 0.25 $\mu_B$ when $J$ goes from 0.0 to 1.0 eV respectively. 
Here again, it is difficult to predict what would be the best $J$ value for NaMnF$_3$ without any experimental feedback.
Previous studies about the DFT$+U$ $J$ parameter report that low $J$ values for $d^5$ orbital filling are usually the best values\cite{bousquet2010b} such as the large values of the $F_z$ canting we obtain at low $J$ (close to 1 $\mu_B$/atom) in the $pPv$ phase of NaMnF$_3$ would be the optimized value.

In order to estimate the error bar of our calculations, we report in Table \ref{tab:Cmcm-all-mag-ord} a list of known magnetic $pPv$ phase in oxide and fluoride perovskites with their related magnetic orderings, $d$ orbital filling and $S$ quantum number.
We note that in all oxide cases as well as in NaNiF$_3$, the ground state spin configuration always corresponds to the low spin state.
According to our calculations, only NaMnF$_3$ has the high spin configuration as ground state.
In order to verify if this is due to the DFT parameters used in our calculations or if this has a physical meaning, we performed calculations of the $pPv$ phase of NaNiF$_3$ and CaRhO$_3$ with $U$=4.0 eV.
We found a low spin ground state in both NaNiF$_3$ ($S$=1) and CaRhO$_3$ ($S$=$\frac{1}{2}$) and so in good agreement with the experimental results.
Going further, we also estimated the non-collinear spin ground state of NaNiF$_3$ and CaRhO$_3$ in their $pPv$ phase.
As expected from the symmetry analysis, we observe canted magnetic ground state in the two systems.
The ground state magnetic ordering of NaNiF$_3$ is found to be $A_z$ with a $G_y$ canting. 
Then, we did not find any weak ferromagnetic spin canting in NaNiF$_3$ in agreement with the experimental data. 
We note that experimentally a transition toward an AFM state with a ferromagnetic canting is observed in the $pPv$ phase of NaNiF$_3$ when applying a small magnetic field.
The direction of the field induced canted FM spins in NaNiF$_3$ are the same than the one we report in the $pPv$ phase of NaMnF$_3$ without magnetic field.
This shows that the high-spin large FM canting we report in NaMnF$_3$ might be metastable in NaNiF$_3$ such as a small magnetic perturbation can switch to this state.
More theoretical and experimental analysis would be welcome to understand this transition.
In the case of CaRhO$_3$ we found $C_y$ with a $F_z$ canting as magnetic ground state. 

No experimental measurements of the amplitude of the spin canting has been reported for $pPv$ CaRhO$_3$ and NaNiF$_3$ but we found that when considering the cases with ferromagnetic spin canting, the canting angle is much smaller in CaRhO$_3$ (4.3$^\circ$) and in NaNiF$_3$ (4.7$^\circ$) than it is in NaMnF$_3$ (11.7$^\circ$, see Table \ref{tab:noncoll-ABX3}).
If we look at the amplitude of the canting per atom we observe a very large amplitude for NaMnF$_3$ (0.95 $\mu_B$/atom) due to the high spin configuration while it is much smaller in the case of NaNiF$_3$ (0.12 $\mu_B$/atom) and of CaRhO$_3$ (0.05 $\mu_B$/atom).
This large canting can be the result of a rather large Dzyaloshinsky-Moriya interaction\cite{Dzyaloshinsky, Moriya} or a large single-ion anisotropy\cite{phys-oxides2004} and more detailed analysis would be required in order to identify the exact origin of this large spin canting.
These calculations allow us to be more confident about the unique high spin ground state of NaMnF$_3$ with a large ferromagnetic spin canting in its $pPv$ phase.

\begin{table}[htbp!]
\caption{Magnetic ordering in fluorides and oxides post-perovskite systems, AFM: antiferromagnetic, DM: diamagnetic, FM: ferromagnetic, CWP: Curie-Weiss paramagnetic. 
*NaNiF$_3$ compound present a magnetic transition from AFM to AFM+FM ordering at applied external magnetic field \cite{Shirako2012}. 
**This work. 
All $pPv$ compounds exhibit a low-spin magnetic configuration but NaMnF$_3$ that according to our calculations is found to be in the high-spin state.}
\centering
\begin{tabularx}{\columnwidth}{X c c c c}
\hline
\hline 
Compound  &   \multicolumn{3}{c}{Magnetic ordering}      &   ref  \rule[-1ex]{0pt}{3.5ex} \\
\hline
             \multicolumn{5}{c}{$pPv$ Oxides $AB$O$_3$}             \rule[-1ex]{0pt}{3.5ex} \\
CaIrO$_3$  &  quasi-1D AFM  &  S=$\frac{1}{2}$ &  Ir$^{4+}$: 5$d^5$(LS)  & \cite{Bogdanov2012}  \rule[-1ex]{0pt}{3.5ex}\\
CaRuO$_3$  &  quasi-1D AFM  &   S=1  &  Ru$^{4+}$: 4$d^4$(LS)  & \cite{Shirako2010, Shirako2011}\rule[-1ex]{0pt}{3.5ex}\\
CaPtO$_3$  &      DM        &   S=0   & Pt$^{4+}$: 5$d^6$(LS)  & \cite{Ohgushi2008, Inaguma2008}\rule[-1ex]{0pt}{3.5ex}\\
CaRhO$_3$  & AFM+weak-FM    &   S=$\frac{1}{2}$ &  Rh$^{4+}$: 4$d^5$(LS)  & \cite{Yamaura2009}  \rule[-1ex]{0pt}{3.5ex}\\
NaIrO$_3$    &      CWP      &  S=1    &      Ir$^{5+}$: 5$d^4$(LS)  & \cite{Nairo2011}  \rule[-1ex]{0pt}{3.5ex}\\
\hline
                \multicolumn{5}{c}{$pPv$ Fluorides $AB$F$_3$}         \rule[-1ex]{0pt}{3.5ex} \\
NaCoF$_3$    &      AFM      &   ---    &      Co$^{2+}$: 3$d^7$(LS)    & \cite{Yusa2012}  \rule[-1ex]{0pt}{3.5ex}\\
NaNiF$_3$    &      AFM*      &   S=1    &      Ni$^{2+}$: 3$d^8$(LS) & \cite{Shirako2012}$^,$**  \rule[-1ex]{0pt}{3.5ex}\\
NaMnF$_3$    &     AFM+FM      &   S=$\frac{5}{2}$    &      Mn$^{2+}$: 3$d^5$(HS) & **  \rule[-1ex]{0pt}{3.5ex}\\
\hline
\end{tabularx}
\label{tab:Cmcm-all-mag-ord}
\end{table}

\begin{table}[htbp!]
\caption{Magnetic orderings allowed in the $Cmcm$ structure for $ABX_3$ systems with $U$ = 4.0 eV and $J$ = 0. 
The large magnetic canting for fluorides compounds can be noted. 
The canting angle ($\varphi$) is measured with respect to the $ab$-plane on the $pPv$ phase. 
For NaNiF$_3$, the two possible magnetic states  \emph{C$_z$F$_y$}  and \emph{C$_y$F$_z$} were computed.}
\centering
\begin{tabularx}{\columnwidth}{c  c  c  c  }
\hline
\hline
Compound & \emph{F$_i$} [$\mu_B$/atom] &  $\varphi$ [deg] & Magnetic Ground-state   \rule[-1ex]{0pt}{3.5ex} \\
\hline
CaRhO$_3$ &  0.05  &   4.3  &  \emph{C$_y$F$_z$}    \rule[-1ex]{0pt}{3.5ex}\\
NaNiF$_3$ &  0.12 (0.15)  &   86.1 (4.7)  &    \emph{C$_z$F$_y$} (\emph{C$_y$F$_z$})   \rule[-1ex]{0pt}{3.5ex}\\
NaMnF$_3$ &  0.95  &   11.7  &   \emph{C$_y$F$_z$}  \rule[-1ex]{0pt}{3.5ex}\\
\hline
\end{tabularx}
\label{tab:noncoll-ABX3}
\end{table}

\begin{figure}[htb!]
 \centering
 \includegraphics[width=7.5cm,keepaspectratio=true]{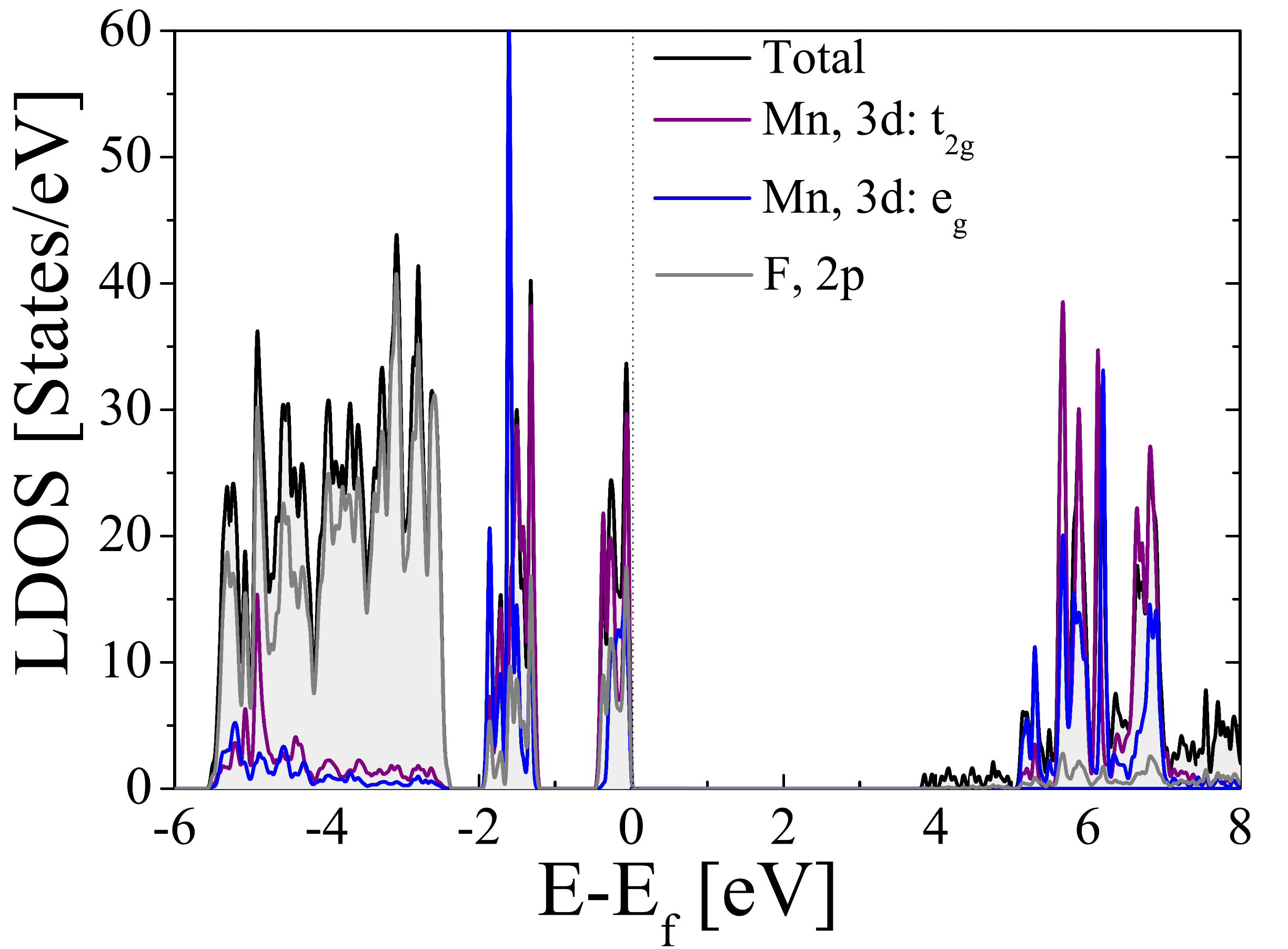}
 \caption{(Color online) Projected local density of states (LDOS) of the \emph{Cmcm} phase of NaMnF$_3$. 
 A regular F-$p$ bonding Mn-$d$ antibonding is observed with small Mn:$3d$ -- F:$2p$ overlaping DOS is responsible for the small superexchange interaction present in NaMnF$_3$.}
 \label{fig:LDOS}
\end{figure}

We also report in Fig.\ref{fig:LDOS}  the local density of states (LDOS) of the $pPv$ phase of NaMnF$_3$.  
The LDOS shows an overlap between the Mn-$3d$ and F-$2p$ states, suggesting a covalent contribution between the Mn and F atoms due to the $p-d$ hybridization, which is a fundamental requirement for the superexchange interaction. 
Moreover, we observe that the fluorine states are lower in energy then the Mn-$d$ orbitals, which corresponds to the expected bonding-antibonding picture of covalency between magnetic cations and anions.
Close to the Fermi level, no contribution of Na states to the DOS is observed such as mostly Na size effects are expected within the structure, a key-point parameter that is at the origin of the geometric ferroelectric instabilities observed in cubic fluoroperovskites.\cite{acgarciacastro2014}
The $d_{xy}$, $d_{xz}$, $d_{yz}$, $d_{z^2}$ and $d_{x^2-y^2}$ orbital occupancy gives the 5 orbitals filled below the Fermi level for a spin channel and the 5 orbitals empty above of the Fermi level for the other spin channel, which confirms the high spin electronic configuration of Mn$^{2+}$ in the $pPv$ phase of NaMnF$_3$.

\subsection*{Conclusions}
We have studied by means of first-principles calculations the high pressure phase diagram of NaMnF$_3$. 
We predicted a phase transition from the $Pnma$ distorted perovskite to a $Cmcm$ post-perovskite phase at a critical pressure of 8 GPa which is relatively low with respect to the similar phase transition observed in other compounds.
As proposed by O'Keeffe\cite{O'keeffe1979} this phase transition fits very well with the octahedra tilting angle requirements where the transition to the $pPv$ is observed when the octahedra tilting angle reach the critical value of 25$^\circ$ with an initial ambient pressure tilting angle of 15$^\circ$.
Unfortunately high pressure experiments have been performed at very high temperature by Akaogi \emph{et. al.}\cite{Akaogi2013} where the O'Keeffe requirement is not fulfilled and also by Katrusiak \emph{et. al.} \cite{Katrusiak1992} who unfortunately did not go beyond 5 GPa such as none of them observed the $pPv$ transition.
Because of the partially filled $d$ orbitals of Mn$^{2+}$ cation we also determined the magnetic ground state of NaMnF$_3$ in its $pPv$ phase.
Our calculations predicted a high spin ground state of the $d^5$ spins with a magnetic moment of 5$\mu_B$ on each Mn.
This high spin structure is particularly interesting and unique since all the other known magnetic compounds presenting the $pPv$ phase present a low spin ground state.
We predicted an AFM magnetic ordering ground state of the $C$ type with the spins lying along the $y$ direction and with a surprisingly large ferromagnetic spin canting along the $z$ direction.
This makes NaMnF$_3$ of particular interest for fundamental and geophysical studies since it has this unique property of high spin magnetic ground state with potentially large ferromagnetic spin canting and because it can be synthesized at relatively small pressure.
The fact that the fluoride postperovskites can be maintained when the pressure is released makes NaMnF$_3$ even more interesting to study.
We hope that our results will motivate the experimentalist to carry out more studies of NaMnF$_3$ in order to confirm the $pPv$ phase transition at low pressure and low temperature and to check its potential for high spin structure with a sizeable canted ferromagnetism. 

\subsection*{Acknowledgements}
This work used the Extreme Science and Engineering Discovery Environment (XSEDE), which is supported by National Science Foundation grant number OCI-1053575 as well as the machines of the Consortium des Equipements de Calcul Intensif (CECI) Funded by F.R.S.-FNRS Belgium.
Additionally, the authors acknowledge the support from the and Texas Advances Computer Center (TACC) and Super Computing System (Mountaineer) at WVU, which are funded in part by the National Science Foundation EPSCoR  Research Infrastructure Improvement Cooperative Agreement 1003907, the state of West Virginia (WVEPSCoR via the Higher Education Policy Commission) and WVU.
A.H. Romero and A.C. Garcia-Castro acknowledge the support of the Marie Curie Actions from the European Union in the international incoming fellowships (grant PIIFR-GA-2011-911070) and CONACyT project 152153.
This work was also supported by F.R.S.-FNRS Belgium (EB).

\bibliography{library}

\end{document}